\documentclass[a4,12pt,epsf, amssymb]{article}
\usepackage{color}
\usepackage{amsmath,amsfonts,latexsym,amssymb}
\usepackage{mathrsfs}

\usepackage[affil-it]{authblk}                         
\usepackage{graphicx,psfrag,epsf,epstopdf}
\usepackage{enumerate}
\usepackage{natbib}
\usepackage{url} 
\usepackage{array, multirow}

\usepackage{moreverb}
\usepackage{lscape}
\usepackage[normalem]{ulem}
\usepackage{bbm}
\usepackage{booktabs}

\newcommand{\bbeta}    {\mbox{\boldmath$\beta$}}
\newcommand{\btheta}   {\mbox{\boldmath$\theta$}}

\newcommand{\balpha}   {\mbox{\boldmath$\alpha$}}

\newcommand{\byi}{\mbox{\boldmath $y$}_{i}}
\newcommand{\bthetat}{\mbox{\boldmath $\theta$}^{(t)}}

\newcommand{\ba}       {{\bf a}}

\newcommand{\bb}       {{\bf b}}

\newcommand{\bc}       {{\bf c}}

\newcommand{\by}       {{\bf y}}

\newcommand{\bz}       {{\bf z}}

\begin{document}


\begin{titlepage}
\clearpage\thispagestyle{empty}

\begin{center}
{\bf Inferring random change point from left-censored longitudinal data by segmented mechanistic nonlinear models, with application in HIV surveillance study}
\\~\\
Hongbin Zhang$^{1,2}$, McKaylee Robertson$^{2}$, Sarah L. Braunstein$^3$, Levi Waldron$^{2,4}$, Denis Nash$^{2,4}$\\
$^1$Department of Biostatistics,
College of Public Health,
University of Kentucky.\\
$^2$Institute for Implementation Science in Population Health,
City University of New York.\\
$^3$Bureau of Hepatitis, HIV, and Sexually Transmitted Infections, Department of Health and Mental Hygiene, New York City.\\
$^4$Department of Epidemiology and Biostatistics, Graduate School of Public Health and Health Policy, City University of New York.

\end{center}

{\small Correspondence: H. Zhang, hongbin.zhang@uky.edu}
\begin{abstract}

The primary goal of public health efforts to control HIV epidemics is to diagnose and treat people with HIV infection as soon as possible after seroconversion. The timing of initiation of antiretroviral therapy (ART) treatment after HIV diagnosis is, therefore, a critical population-level indicator that can be used to measure the effectiveness of public health programs and policies at local and national levels. However, population-based data on ART initiation are unavailable because ART initiation and prescription are typically measured indirectly by public health departments (e.g., with viral suppression as a proxy). In this paper, we present a random change-point model to infer the time of ART initiation utilizing routinely reported individual-level HIV viral load from an HIV surveillance system. To deal with the left-censoring and the nonlinear trajectory of viral load data, we formulate a flexible segmented nonlinear mixed effects model and propose a Stochastic version of EM (StEM) algorithm, coupled with a Gibbs sampler for the inference. We apply the method to a random subset of HIV surveillance data to infer the timing of ART initiation since diagnosis and to gain additional insights into the viral load dynamics. Simulation studies are also performed to evaluate the properties of the proposed method.

\end{abstract}

\noindent{\it Key words}: HIV surveillance, Random change-point model, Nonlinear mixed effects model, Censored data, Stochastic version of EM, Gibbs sampler, Antiretroviral therapy initiation.

\end{titlepage}
\vfill

\newpage
\section{Introduction}
\label{sec:intro}
Antiretroviral therapy (ART) reduces the level of HIV viral load (VL), resulting in improved health outcomes for the individuals and reduced risk of HIV transmission. It is estimated that if a ``test and treat" policy is widely and effectively implemented, HIV incidence and prevalence could be significantly reduced over the coming decades \citep{granich2009}. However,
the time from HIV diagnosis to ART initiation is a critical determinant of success and
is not uniform across jurisdictions and patients due to various barriers and other factors. While time-updated laboratory testing data on viral loads (RNA copies per milliliter of blood plasma) are reported in HIV case surveillance activities, information on ART is not collected. It is not feasible to get time-updated clinical information following HIV diagnosis from HIV care providers as part of routine surveillance. ART initiation is typically measured indirectly by public health departments using viral suppression as a proxy or via clinical cohorts.  \cite{braunstein2016} developed the first case surveillance-based measure of HIV treatment initiation utilizing routinely available HIV viral load testing results to infer the timing of ART initiation among persons newly diagnosed with HIV. The study is empirical, e.g., using the middle of a time interval as ART initiation is not justified.   Motivated by the study, this paper considers a statistical modeling method, e.g., using a change-point model to estimate and infer ART initiation time. A change-point model has a biological interpretation as the ART initiation induces trajectory change of the underlying process of viral loads. The analysis is complex due to data truncation issues when viral loads under a certain threshold are `undetectable' due to the lower quantification limit of the VL assay. In this paper, we aimed to formulate a flexible change-point model to describe the trajectory of viral loads and estimate the time point of ART initiation.

Random change-point models allowing individual-specific changes for longitudinal outcomes have been widely used in medical research \citep{dominicus2008}. A general way to define a class of random change-point models is to assume a linear mixed effects model for the longitudinal data before and after the change point, i.e., segmented linear mixed effects model \citetext{see, e.g., \citealp{rudoy2010}, \citealp{moss2016}, \citealp{Buhule2020}}. In some applications such as ours above, however, the longitudinal data may be censored, representing measurements from bio-assays where the exact quantification below a certain threshold is impossible. In such a case, the assumed linear model based on the observed
longitudinal data may be inappropriate for the (unobserved) data. If, on the
other hand, a mechanical or scientific model is available for the longitudinal data, such
a mechanical model can be used to better ``predict" the unobserved data, leading to a better estimate of the change points. Such a mechanical model is often nonlinear. 
In this article, we consider random change-point
models where we model the longitudinal data subject to left censoring due to detection limits by nonlinear mixed effects (NLME) models.

To our knowledge, the random change-point model with censoring has not been studied in the literature. For censored response without change point, \cite{wul2002} considered an approximated likelihood method with the NLME model. For censored covariate, a joint modeling approach was proposed by \cite{zhang2018sim} and \cite{zhang2018jrss} using NLME model for the covariate with discrete and survival outcome, respectively, implemented by Monte Carlo Expectation-Maximization (MCEM) algorithm. Regarding the methods around random change-point models, most implementations in statistical literature have been from a Bayesian perspective. The major challenge of accounting for a random change point in a likelihood framework is computational since closed form expressions are often not available \citetext{see, e.g., \citealp{naumova2001}, \citealp{hall2003}, and \citealp{muggeo2014}}. In our case, it is further complicated by the nonlinear models and the data censoring. 
 
This paper aims to propose a segmented NLME model
for left-censored data and consider a full likelihood-based inference via the stochastic version of the EM algorithm, a.k.a. StEM algorithm,
proposed by \cite{diebolt1996}.
The StEM algorithm is more computationally efficient
than the MCEM algorithm as only one realization of the missing data is required for each iteration \citep{ip2002}. Most recently, \cite{wang2020} extended the StEM algorithm to estimate parameters in VL dynamics models accounting for left-censored data; the authors showed that the resulting
estimator is less biased than naive methods that either omit all censored data points or impute the censored observation
with half of the quantification limit.

In what follows, we describe an HIV surveillance registry that motivated our research, the VL dynamics, and the nonlinear random change-point model in Section 2. In Section 3, we present the model in general form and describe the likelihoods and the estimation procedure.  In Section 4, we analyze a randomly selected subset of data from the HIV surveillance registry. We evaluate the method with simulation in Section 5. We conclude the article with some discussions in Section 6. 

\section{HIV Surveillance Registry and Nonlinear Random Change-Point Model for VL Dynamics}

\subsection{The HIV Surveillance Registry and VL Dynamics}

New York City (NYC) is a major epicenter of the HIV epidemic. New York State public health law requires named reporting of all  positive HIV
diagnostic tests to the NYC Department of Health
and Mental Hygiene (DOHMH). The DOHMH's HIV Epidemiology Program maintains the
population-based HIV surveillance registry, which is continuously
updated with demographic information on persons
meeting the Centers for Disease Control and Prevention's HIV surveillance case definitions and
with results of all laboratory tests such as viral loads and CD4 counts conducted in
NYC. This paper's target population of inference includes persons newly diagnosed with HIV between 2006 and 2015 and aged $\geq$ 13 years at HIV diagnosis and their laboratory information reported through December 31, 2017. The population level data contain biomarkers information and corresponding reporting time points from 27,639 persons with HIV.

We aim to infer the timing of treatment initiation among those individuals utilizing the serial HIV viral load measurements. In the absence of ART, viral load shows a dramatic fluctuation after HIV infection before reaching a set point. It will then increase steadily until the development of AIDS if without treatment \citep{mei2008}. ART initiation, however, induces substantial reductions in HIV viral load. In the HIV registry, less than 6\% of individuals' diagnosis times are known to be in the acute phase of HIV infection. To simplify the modeling, we assume that the HIV diagnosis occurs after the set-point and focus on modeling VL dynamics that the ART initiation might alter after the diagnosis.

\subsection{Nonlinear Random Change-Point Model for VL Dynamics}

The focus of longitudinal data analysis is typically on time trajectory to investigate how a longitudinal outcome evolves in time. Random effects change-point models enable the analysis to account for the change in the time trajectory by including individual change points. The change is usually induced by external events, so the data deviate from their original course. The conventional random effects change-point models postulate that the longitudinal outcome is constructed with segmented linear mixed effect models. However, as sketched in the Introduction, linearity assumptions may be restrictive in many applications. Even though linear models may fit the observed data well, they might be inappropriate for data subject to censoring, as occurs with HIV viral loads in HIV data, often after the ART treatment. There is extensive research on VL dynamics after ART, which assess antiretroviral drug's therapeutic effect. Based on biological and clinical arguments, \cite{wd1999} proposed to approximate viral load data pattern by the viological model $ V(t) = P_{1} + P_{2} e^{-\lambda}$ where $V(t)$ is the total virus at time $t$ and $P_{1}$ and $P_{2}$ are baseline values. Parameters\ $\lambda$ is the viral decay rates and may be interpreted as the turnover rates of productively infected cells and long-lived or latently infected cells if the therapy is perfect. Detailed discussion of the model has been given by \cite{grossman1999} and \cite{perelson1996}.

For the problem at hand, we may consider the following mixed effects model for the reported viral loads for individual $i$ at time $t_{ij}$ after diagnosis in HIV surveillance data
\[
\begin{split}
&\log_{10}(V(t_{ij})) = a_{1i} (t_{ij} - \tau_{i})^{-} + \log_{10}(b_{1i} + b_{2i}e^{-b_{3i}(t_{ij}-\tau_{i})^{+}}) + e_{ij},\\
\end{split}
\]
\noindent where the $\log_{10}$ transformation is used to stabilize the variance and makes the data more normally distributed. The error term is represented by $e_{ij}$, $\tau_{i}$ is the change point, functions $x^{-}$ and $x^{+}$ correspond to $\text{min}(x, 0)$ and $\text{max}(x, 0)$, respectively. The quantity $a_{1i}$ is the subject-specific regression coefficient representing the slope of viral loads before the change point, and $b_{1i}$,  $b_{2i}$, $b_{3i}$ are subject-specific mixed effects for the viral trajectory after the change point. We may define
\[
\begin{split}
 a_{1i} &= \alpha_{1} + \alpha_{1i}, \quad b_{1i} = \beta_{1} + \beta_{1i}, \quad b_{2i} = \beta_{2} + \beta_{2i},  \quad b_{3i} = \beta_{3} + \beta_{3i}. \\
\end{split}
\]
\noindent where $\alpha_{1}$, $\beta_{1}$, $\beta_{2}$, and $\beta_{3}$ are the population parameters (fixed effects), $\alpha_{1i}$, $\beta_{1i}$, $\beta_{2i}$, and $\beta_{3i}$ are the random effects which are usually assumed to follow normal distribution with zero mean.  

The choice of the distribution for the random change points is a model assumption and will depend on the process under investigation. For example, the distribution may be conventionally assumed to be normal. A log-normal distribution may be considered for the type of change point with a positive value. Choosing a parametric population distribution for the change points makes it possible to pool information: instead of estimating change points individually, their distribution parameters are estimated. In addition, when a change point is modeled as a random effect, the so-called problem of first-order discontinuity \citep{tishler1981} disappears since the parameters of the change-point distribution are estimated instead of a fixed-effect change point.

\section{An Estimation Procedure Based on StEM Algorithm}

\subsection{The Models and Notations}

In this section we present the models and the methods in general forms, illustrating that our methods may be applicable in other applications. Let $y_{ij}$, $j = 1, 2, \ldots, n_{i}$, be the measurement (can be left-censored) for subject $i = 1, 2, \ldots, n$ at time $t_{ij}$. We consider a general segmented-NLME model
\begin{equation}
\begin{split}
 y_{ij} &= g((t_{ij}-\tau_{i})^{-}, \ba_{i}) + h((t_{ij} - \tau_{i})^{+}, \bb_{i}) + e_{ij},  \\
  \tau_{i} &\sim N(\tau, \sigma_{\tau}^2), \quad \ba_{i} \sim N(\balpha, A), \quad \bb_{i} \sim N(\bbeta, B), \quad e_{ij}|\tau_{i}, \ba_{i}, \bb_{i} \sim N(0, \sigma_{e}^{2}), \\
 \end{split}
\end{equation}
\noindent where $g(\cdot)$ and $h(\cdot)$ are known nonlinear functions, $\balpha$ and $\bbeta$ are vectors of population parameters, $\tau$ is the population mean and $\sigma_{\tau}^2$ is the variance for the random change point $\tau_{i}$, $A$ and $B$ are the variance-covariance matrix for random effects $\ba_{i}$ and $\bb_{i}$, respectively, and $\sigma_{e}^2$ is the within-individual variance. Function $x(\cdot)^{-}$ and $x(\cdot)^{+}$ are defined the same as in the last Section. For the segmented-NLME model, it is reasonable to assume that $\ba_{i}$, $\bb_{i}$ are independent and both are independent of $\tau_{i}$ which is usually introduced externally.

We consider a likelihood-based estimation and inference  procedure for the model (3.1)
using the observed data $\{ (\by_{i}, \bc_{i}), i=1,\ldots, n \}$ where $\bc_{i}=(c_{i_{1}},\ldots, c_{i_{ni}})$ is the vector of censoring indication with $c_{ij}=1$ when $y_{ij}$ is left-censored and 0 otherwise. Let $\btheta=( \balpha, \bbeta, \tau, \sigma_{e}^2, \sigma_{\tau}^2,  A, B)$ be the collection of all unknown parameters and $f(\cdot)$ be a generic density function, and let $f(X|Y)$ denote a conditional density of $X$ given $Y$. The observed data likelihood can be written as
\begin{equation}
\begin{split}
 L(\btheta) & =  \prod_{i=1}^{n} \Big \{\int \int \int \Big [ \prod_{j=1}^{n_{i}}
f(y_{ij}|\tau_{i}, \ba_{i}, \bb_{i}; \btheta)^{1-c_{ij}}F(d|\tau_{i}, \ba_{i}, \bb_{i}; \btheta)^{c_{ij}}  \Big ] \\  
& \quad \quad \quad \quad \quad \quad \quad \quad \quad \quad \quad \quad \quad \quad \quad \quad \quad \times f(\tau_{i})f(\ba_{i})f(\bb_{i}) d\tau_{i}  d\ba_{i} d\bb_{i} \Big \},\\
  \end{split}
\end{equation}
\noindent where $d$ is the detection limit and
\[
F(d|\tau_{i}, \ba_{i}, \bb_{i}; \btheta) = \int_{-\infty}^{d} f(y_{ij}|\tau_{i}, \ba_{i}, \bb_{i}; \btheta) dy_{ij}.
\]
Directly maximizing the likelihood (3.2) is challenging due to the nonlinear models involved and the nested integrals. Writing $\byi=(\by_{obs,i}, \by_{cen,i})$, by treating $\by_{cen,i}$ (the censored component of $\by_{i}$) and the unobserved random effects $\tau_{i}$, $\ba_{i}$, $\bb_{i}$ as ``missing data", we have ``complete data" $\big \{ (\byi, \tau_{i}, \ba_{i}, \bb_{i}), i=1,\ldots, n \big \}$ and the complete-data log-likelihood function for individual $i$ can be expressed as
\begin{equation}
\begin{split}
l_{c}(\btheta)  = \log f(\tau_{i}; \btheta) + \log f(\ba_{i}; A) + \log f(\bb_{i}; B) + \log f(\by_{i}|\tau_{i}, \ba_{i}, \bb_{i};  \btheta).\\
\end{split}
\end{equation}

\subsection{The Estimation Procedure}

The EM algorithm introduced by \cite{dempster1977} is a classical approach to estimate parameters of models with non-observed or incomplete data. Let us briefly cover the principle. Denote $\bz$ as the vector of non-observed data, $(\by, \bz)$ the complete data and $L_{c}(\by,\bz;\btheta)$ the log-likelihood of the complete data, the EM algorithm maximizes the $Q(\btheta|\btheta')=E(L_{c}(\by,\bz;\btheta)|\by;\btheta')$ function in two steps. At the ${k}^{th}$ iteration, the E-step is the evaluation of $Q^{(k)}(\btheta) = Q(\btheta|{\btheta}^{(k-1)})$, where the M-step updates ${\btheta}^{(k-1)}$ by maximizing $Q^{(k)}(\btheta)$.

For the cases in which the E-step has no analytic form, \cite{wt1990} introduce the Monte Carlo EM (MCEM) algorithm, which calculates the conditional expectations at the E-step via many simulations within each iteration and hence is quite computationally intensive. \cite{diebolt1996} introduce stochastic versions of the EM algorithm, namely the stochastic EM (StEM), which replaces the E-step with a single imputation of the complete data and then averages the last batch of $M$
estimates in the Markov Chain iterative sequence to obtain the point estimate of the parameters.  The imputed data $\bz^{(k)}$ at the $k^{th}$ iteration are a random draw from the conditional distribution
of the missing data given the observed data and the estimated parameter values at the $(k-1)^{th}$ iteration,
$f(\bz^{(k)}|\by, \btheta^{(k-1)})$. As $\bz^{(k)}$ only depends on $\bz^{(k-1)}$, $\{\bz^{(k)}\}_{k \geq 1}$ is a Markov chain. Assuming that $\bz^{(k)}$ take values in a compact space and the kernel of the Markov chain
is positive continuous for a Lebesgue measure, the Markov chain is ergodic, and that ensures the existence of a unique stationary distribution \citetext{see, e.g., \citealp{ip1994}, \citealp{nielsen2000}}.

We now detail the StEM algorithm for the segmented-NLME previously presented. At the $(k+1)^{th}$ iteration:

{\bf Imputation:} Draw missing data $(\tau_{i}, \ba_{i}, \bb_{i}, \by_{cen,i})$ from the conditional distribution $[\tau_{i}, \ba_{i}, \bb_{i}, \by_{cen,i}|\by_{obs, i}; \btheta^{(k)}]$. Specifically, we use the Gibbs sampler to generate samples from $[\tau_{i},\ba_{i}, \bb_{i}, \by_{cen,i}|\by_{obs,i}; \btheta^{(k)}]$ by iteratively sampling from the full conditionals $[\tau_{i}|\byi,\ba_{i}, \bb_{i};\btheta^{(k)}]$, $[\ba_{i}|\byi,\tau_{i}, \bb_{i};\btheta^{(k)}]$, $[\bb_{i}|\byi,\tau_{i},\ba_{i};\btheta^{(k)}]$, and $[\by_{cen,i}|\by_{obs,i}, \tau_{i}, \ba_{i}, \bb_{i}; \btheta^{(k)}]$, as follows:
\[
\begin{split}
 &f(\tau_{i}|\by_{i},\ba_{i}, \bb_{i};\btheta^{(k)}) \propto f(\tau_{i}; \btheta^{(k)})f(\by_{i}|\tau_{i},\ba_{i},\bb_{i};\btheta^{(k)}),\\
 &f(\ba_{i}|\by_{i},\tau_{i},\bb_{i};\btheta^{(k)}) \propto f(\ba_{i};\btheta^{(k)})f(\by_{i}|\tau_{i},\ba_{i},\bb_{i};\btheta^{(k)}),\\
 &f(\bb_{i}|\by_{i},\tau_{i},\ba_{i};\btheta^{(k)}) \propto f(\bb_{i};\btheta^{(k)})f(\by_{i}|\tau_{i},\ba_{i},\bb_{i};\btheta^{(k)}),\\
 &f(\by_{cen,i}|\by_{obs,i}, \tau_{i}, \ba_{i}, \bb_{i}; \btheta^{(k)}) \propto f(\by_{i}|\tau_{i},\ba_{i},\bb_{i}; \btheta^{(k)}).\\
 \end{split}
\]

Monte Carlo samples from each of the full conditionals can be obtained through multivariate rejection sampling methods (see Appendix). 

{\bf Maximization:} With data augmentation, the maximization step involves maximizing the complete likelihood (3.3). For the ``complete data" $\big \{ (\byi, \tau_{i}, \ba_{i}, \bb_{i}), i=1,\ldots, n \big \}$, the complete log-likelihood can be written as summation of four parts
\[
 L_{c}(\btheta) = L_{c}^{(1)}(\btheta) + L_{c}^{(2)}(\btheta) + L_{c}^{(3)}(\btheta) + L_{c}^{(4)}(\btheta)
\]
where
\[
\begin{split}
 L_{c}^{(1)}(\btheta) &= \sum_{i=1}^{n} \log \Big [\sigma_{\tau}^{-1} \exp \Big (-\frac{(\tau_{i} - \tau)^2}{2\sigma_{\tau}}  \Big ) \Big ], \\
 L_{c}^{(2)}(\btheta) &= -n\log (2\pi) - \frac{n}{2}\log |A| - \frac{1}{2} \sum_{i=1}^{n} (\ba_{i} - \balpha)^{T} A^{-1} (\ba_{i} - \balpha) ,\\
 L_{c}^{(3)}(\btheta) &= -n\log (2\pi) - \frac{n}{2}\log |B| - \frac{1}{2} \sum_{i=1}^{n} (\bb_{i} - \bbeta)^{T} B^{-1} (\bb_{i} - \bbeta),\\
 L_{c}^{(4)}(\btheta) &= \sum_{i=1}^{n} \log \Big \{ \prod_{j=1}^{n_{i}} \Big [\sigma_{e}^{-1} \exp \Big (-\frac{(y_{ij} - g((t_{ij} - \tau_{i})^{-},\ba_{i}) - h((t_{ij}-\tau_{i})^{+}, \bb_{i}))^2}{2\sigma_{e}^2} \Big ) \Big ]   \Big \}.\\
\end{split}
\]

Since $(\byi, \tau_{i}, \ba_{i}, \bb_{i})$ are regarding as data, the complete log-likelihood no longer involves integrals, which substantially simplifies the maximization. Also, due to the mutual independence among $\tau_{i}$, $\ba_{i}$, $\bb_{i}$, the maximization can be done by part. Solving the score equations yields the following estimations:

\[
\begin{split}
 \tau &= \frac{1}{n}\sum_{i=1}^{n}\tau_{i}, \quad \sigma_{\tau}^2 = \frac{1}{n} \sum_{i=1}^{n} (\tau_{i} - \tau)^2, \\
 \balpha &= \frac{1}{n}\sum_{i=1}^{n}\ba_{i}, \quad A = \frac{1}{n}\sum_{i=1}^{n}(\ba_{i}-\balpha)(\ba_{i}-\balpha)^{T}, \\
 \bbeta &= \frac{1}{n}\sum_{i=1}^{n}\bb_{i}, \quad B = \frac{1}{n}\sum_{i=1}^{n}(\bb_{i}-\bbeta)(\bb_{i}-\bbeta)^{T}, \\
 \sigma_{e}^2 &= \frac{1}{n}\sum_{i=1}^{n} \Big \{\frac{1}{n_{i}} \sum_{j=1}^{n_{i}} \Big [y_{ij} - g((t_{ij} - \tau_{i})^{-},\ba_{i}) - h((t_{ij}-\tau_{i})^{+}, \bb_{i})\Big ]^2 \Big \}. \\
\end{split}
\]

Estimates of standard
errors can be obtained as the inverse of the Fisher information matrix \citetext{see for instance \citealp{walter2007}}.
As with the likelihood defined in (3.2), the Fisher information matrix of the segmented-NLME has no closed form solution. Approximations to the Fisher information matrix
have been proposed in the optimal design context by Mentré and others \citetext{\citealp{mentre1997}; \citealp{retout2007}}. In this paper, we compute the Fisher information
matrix by linearizing of the function $g(\cdot) + h(\cdot)$ around the conditional expectation of the individual
Gaussian parameters \{$\tau_{i}, \ba_{i}, \bb_{i}, i=1,\ldots,n \}$ with imputed $\by_{i}$. The resulting model is Gaussian, and its Fisher
information matrix is a block matrix (no correlations between the estimated fixed effects
and the estimated variances). Alternatively, the
Louis principle could be used to compute the Fisher information matrix based on the complete data likelihood where the second derivatives of the likelihood are involved \citep{louis1982}.
\vspace{0.5cm}

\section{Data Analysis}
\label{sec:data}
We present the analysis of the HIV surveillance registry data introduced in Section 2. We aim to estimate the ART initiation time using the routinely reported viral loads from the HIV surveillance registry. Due to computing resource constraints, we consider a random subset of 500 individuals with at least two viral reports. The random sample enables the change point inference for the corresponding target population. 

To obtain the initial values of the parameters, we start with a modified version of the {\sl log1plus} algorithm from \cite{braunstein2016}, denoted as {\sl log1plus$^{*}$}. Same as the original {\sl log1plus} algorithm, the {\sl log1plus$^{*}$} algorithm detects ART occurrence by sequentially examining pairs of reported viral loads over time for an individual. ART initiation is detected if the difference between the two viral loads drops more than one log10-based unit within an observing window with specific width (e.g., three months) or the viral load measures change from detectable to undetectable (i.e., left-censored). While the original {\sl log1plus} was used to detect ART initiation and make reference to the detected sub-population, as a preliminary step to obtain initial values for the interested parameters, the {\sl log1plus$^{*}$} extends the algorithm to the entire sample. Also, instead of using the middle point as ART initiation time, {\sl log1plus$^{*}$} used the first reporting time of the VL pair upon which ART occurrence is detected. Even though the approach is still empirical, such a choice of ART initiation is more biologically plausible since the VL is likely to decline right after the ART. For those individuals with no ART detected, a random time point beyond the last VL reporting time was used as the initial change point. 

 Due to drug resistance, inadequate drug absorption, or sub-optimal medication adherence, a viral rebound can occur \citep{murray1999}. To include many relevant data as possible and to simplify the modeling, we preserve any reports before the first VL reporting time of the VL pair which detects the ART and extend time points (and corresponding VL values) after the second reporting time of the same VL pair until viral rebound. Viral rebound is defined in this analysis as the occurrence of a larger viral load value (comparing to the previous measure) or when two adjacent VL measures change from undetectable to detectable over time. The final analytic data have a total of 2621 viral reports with reporting period ranging from 0 to 3 years after HIV diagnosis and reporting frequency ranging from 2 to 16, with a median of 5 for the 500 random samples. Thirty-seven percent of viral loads were under the detection limits. 

We fit several candidate models upon the viral loads separated by the initial change points to obtain the starting values for the parameters in the segmented models. For the pre- ART segment, we choose from a linear mixed effects model with linear or quadratic term of the time variables. For the post- ART segment, there are the one-compartment and two-compartment nonlinear mixed effects \citep{wd1999} to select. Based on AIC, we decide on a combination of a linear mixed effects model with linear time term and a bi-exponential nonlinear mixed effects model to implement the StEM algorithm. Specifically, the segmented-NLME model for the viral measure $y_{ij}$ at time $t_{ij}$ is: 
\begin{equation}
\begin{split}
&y_{ij} = g((t_{ij}-\tau_{i})^{-}, \ba_{i}) + h((t_{ij} - \tau_{i})^{+}, \bb_{i}) + e_{ij},  \\
&g((t_{ij}-\tau_{i})^{-}, \ba_{i}) = a_{i} (t_{ij} - \tau_{i})^{-},\\
&h((t_{ij}-\tau_{i})^{+}, \bb_{i}) = \log_{10}(b_{1i}e^{-b_{2i}(t_{ij} - \tau_{i})^{+}} + b_{3i}e^{-b_{4i}(t_{ij}-\tau_{i})^{+}}).\\
\end{split}
\end{equation}

The model has a six-dimensional structure of random effects: $a_{i}$ is for the slope of the pre- change point viral loads, $\tau_{i}$ is the change point, $b_{1i}$, $b_{3i}$ are the base values and $b_{2i}$, $b_{4i}$ are the decay rates for the 2-phases post- change point viral loads. While there is no constraint on the range of $a_{i}$, other random effects are biologically positive. We therefore use normal distribution for $a_{i}$ and log-normal distribution for other random effects, e.g., we assume $a_{i} \sim N(\alpha, \sigma_{A}^2)$, $\tau_{i} \sim LN(\tau, \sigma_{\tau}^2)$ $\bb_{i} \equiv  (b_{1i}, b_{2i}, b_{3i}, b_{4i})^{T} \sim LN ((\beta_{1}, \beta_{2}, \beta_{3}, \beta_{4})^{T},  B)$ where the variance-covariance matrix $B$ for the random effects $\bb_{i}$ is assumed to be unstructured to allow for all possible correlations. 



Determining convergence appears to be an open question for StEM. In the literature, the commonly used approach for convergence diagnostics is through visual examination of the trace plots \citetext{see, e.g., \citealp{yang2018}, \citealp{wang2020}, \citealp{huang2020}}. Recently, \cite{zhang2020} proposed a Geweke Statistics based method. We adopt this approach in our implementation and claim the convergence is achieved when the Geweke statistics is smaller than a designated threshold (see Simulation Section for details). With our data and method, the convergence can usually arrive before 2000 iterations. We use statistical software R for the entire implementation.

Figure 1 shows the trace plots of each parameter in model (4.4). A total of 2000 iterations are displayed where the convergence is arrived at the 1940$^{th}$ iteration. We use the estimates from the last 300 iterations (i.e., $M=300$) to calculate the final point estimate and the standard error as 
\[
\hat{\theta} = \frac{1}{M}\sum_{m=1}^{M}\hat{\theta}^{(m)}, \quad \hat{\sigma}_{\hat{\theta}} = \sqrt{\frac{1}{M}\sum_{m=1}^{M} \hat{\sigma}_{\hat{\theta}^{(m)}}^{2} + (1 + \frac{1}{M})\frac{1}{M-1} \sum_{m=1}^{M} (\hat{\theta}^{(m)} - \hat{\theta})^{2} },
\]
where $\hat{\sigma}_{\hat{\theta}^{(m)}}$ is obtained from the linearization method described in Section 3. 

Table 1 shows the estimations for the sample of HIV surveillance registry data from the StEM algorithm and from the {\sl $log1plu^{*}$} algorithm. Figure 2(a) shows the population level viral trajectory estimated by the random change-point model for the first two years after HIV diagnosis. The pre-ART trend of viral dynamic is quite flat but increasing where a positive slope is estimated as $\alpha=0.14$ (se=0.03). With ART, the viral load trend shows a sharp drop in the first phase, followed by a slow decay, i.e., the second phase. The population mean parameter $\tau$ for the change point (ART initiation time) is estimated as -1.15, (se=0.17) which  corresponds to an inverse log transformed scale of 116 days after HIV diagnosis, with a 95\% confidence interval from 83 to 161 days (as contrast, the empirical method yields an estimate of the change point as -1.32 with se=0.11, which is corresponding to 98 days, with 95\% confidence interval from 79 to 121 days). The distribution of the change points is characterized by both $\tau$ and $\sigma_{\tau}^2$. The density plot of this distribution can be seen in Figure 2(b) where both StEM estimation and the empirical distribution of the ART initiation time from the {\sl log1plus$^{*}$} algorithm are displayed. The distributions are right skewed while the StEM method tends to estimate larger (later) ART initiation time than the one from the {\sl log1plus$^{*}$}. The {\sl log1plus$^{*}$} based change-point distribution is heavily influenced by the empirical approach of picking the ART initiation time, e.g., from the first reporting time of the VL pair that detects the ART. Also, the distribution is not smooth where we see a slight upward bump, e.g., around 1.5 year after HIV diagnosis, reflecting the empirical determination of the change points (e.g., as a random value beyond the last VL reporting time) for those individuals without ART.   

We also perform model diagnostics by comparing the observed values with fitted values for each individual in the dataset. We predict the individual random effects, including the change point, by the posterior mean obtained from 100 additional Monte Carlo samples of the full conditionals at the convergence. Figure~\ref{fig:Fig3} shows the fitted versus the observed viral loads from nine individuals who are chosen to represent different patterns. The predicted change point and corresponding viral load at ART treatment time are marked in the plot. Estimated population level viral trajectory is also displayed. The two phases of viral decay can be seen, varying by individual. The segmented-NLME model predicts the left-censored viral loads to follow such a trend beneath the detection limit (see id = 2, 5, 6, and 8). Although restricted to be positive by the model, we see that our model predicts close to zero (HIV diagnosis time) ART initiation time, e.g., for the three individuals (id = 1, 2, and 3) where a sharp decline of viral load after the change point is observed. Compared to the initial change-point value picked by the {\sl log1plus$^{*}$}, the StEM algorithm can predict quite different ART initiation times, as seen from individuals with id 4 and 7. The StEM-based estimations are influenced by 1) the population parameters, 2) each individual's observed viral loads, and 3) the assumed model.

Interestingly, the StEM can predict an ART initiation time before the first virus reporting time, as seen in the case for id=8 where a relatively low value (less than 2.5, log10 transformed scale) is seen at the first viral reporting time. The real-world interpretation for such cases might be that the individual received ART without a VL test on the same day. A situation like this is atypical but plausible, and we present such a case to show the model-based estimation's resilience. For example, the predicted viral load at the ART initiation time seems reasonable. The last case, id 9, represents the situation in which ART initiation is predicted beyond the time frame of available viral reports. 

\section{Simulation}

We evaluate the performance of the StEM algorithm through simulation. We generate data to mimic the viral reports in the HIV surveillance registry and present results for the settings with and without left-censored values. Each simulated dataset contains 500 individuals as in the sample for the real data analysis. We emulate the irregular reporting time since HIV diagnosis by a progressive state-transition model with first-order Markov assumption, i.e., the length of lagging for a reporting time depends on the previous reporting time. We therefore generate the stochastic measurement time $(t_{i1}, t_{i2}, \ldots, t_{in_{i}})$ based on parameters from the fitted model on the actual reporting time points in the HIV surveillance registry data. Specifically, the viral load reporting time $T$ is assumed to follow an exponential distribution with parameter $\xi > 0$. Given the previous reporting time $u >0$, the next reporting time, conditioning on $u$, is
\[
 T_{|u} = \exp \Big( \log(-\frac{1}{\xi}\log(X) + u) \Big ),
\]
for $X \sim \text{Uniform}(0,1)$. In the simulation, we set $\xi=1.45$ which is the estimate from the real data. We then generate viral loads using model (4.4). The values for the fixed effects $(\alpha, \beta_{1}, \beta_{2}, \beta_{3}, \beta_{4}, \tau)$, the corresponding variance components $(\sigma_{A}^2, B, \sigma_{\tau}^2)$, and the variance of error $\sigma_{e}^2$ are provided in Table 1 (StEM based results). The left-censoring is simulated by randomly setting a portion (0\%, 30\% and 50\%) of the generated viral loads that are below 200 RNA copies per milliliter of blood plasma, in the original scale of the biomarker.

We conduct 100 simulation runs for each scenario and with each run, once started the Markov chain, as described in the section above with the initial values, the stationarity is determined by using a batch procedure based
on the Geweke statistic \citep{gelman1992}. Let batch size be $M$, where $M$ is set to 300 as the default value in real data analysis and simulation. We use a moving
window for the Markov chain and compute the Geweke statistics at each increment of 10 iterations ($w$=10). More precisely,  the batch procedure goes in the following steps:

1. {\bf Initialization}. Set $B=0$. Run the StEM algorithm to obtain the initial series of the estimates $\{\btheta^{(w*B+1)}, \ldots, \btheta^{(w*B+M)}\}$.

2. {\bf Check stationarity}. For each entry $p$ in $\btheta$, we compute the Geweke statistic $z_{p}$ from the Markov chain $\{\theta_{p}^{(w*B+1)}, \ldots, \theta_{p}^{(w*B+M)}\}$, based on the standardized mean difference between first 10\% and last 50\% part of the chain. We regard stationary being reached when all $|z_{p}|s$ are sufficiently small, i.e.,
\[
    \sum_{p=1}^{P}z_{p}^2 < \epsilon P,
\]
where P is the total number of parameters and $\epsilon$ is set to be 1.5 in the implementation as in \cite{zhang2020}.

3. {\bf Update.} If stationarity is not reached, execute $w$ additional runs of the chain, increase the number B by 1 and then repeat step 2.

At convergence over the simulation replicates, besides the average standard error (SE) and standard deviation (SD) of the multiple estimates of the parameter, we also calculate the mean squared error (MSE) and the percentage of bias (Bias\%) by comparing the estimate of the parameter with the true value (see Table 2 for the definitions). Table 2 presents the simulation results for the fixed effects, where we also included the results from the {\sl log1plus$^{*}$} based method for comparison. We see the SEs and SDs are generally agreeable, indicating both methods work well for the finite sample. When there is no left-censoring, the {\sl log1plus$^{*}$} method is comparable with the StEM method. With the censoring, the StEM-based procedure produces substantially smaller MSE. It has much less bias, especially regarding the estimation of $\beta_{4}$ (the population parameter for the second phase decay rate of VL) and $\tau$. As the censored amount increases, the  {\sl log1plus$^{*}$} method shows more deteriorated performance while the deterioration with the StEM method is barely noticeable.

\section{Conclusion and Discussion}
Nonlinear models have broad applications in HIV studies, cancer research, and pharmacokinetic modeling \citep{lindsey2001}. In this paper, we extend the random change-point model to a more general class, allowing for nonlinear mixed effects models for each segment. In addition, we establish a StEM-based solution for the left-censoring problem that occurred in the longitudinal data and evaluate the convergence criteria of the Markov chain through simulation. The proposed method has conceptual simplicity, attributing to the EM algorithm, providing a maximum likelihood solution beyond the Bayesian framework.  

When the method is applied to the data, we effectively extend our previous study in several important directions. For example, using a randomly selected sub-sample, we provide a model-based estimation of ART initiation time after HIV diagnosis for the target population. Simulation shows such a model-based metric is more accurate than the empirical version. Furthermore, regarding the viral load dynamics, it is generally hypothesized that the severity of HIV infection would increase without treatment, leading to a higher risk of contracting AIDS. With the segmented-NLME model, we further confirm such a hypothesis: an estimated positive slope of the pri- ART initiation trend for the population. In addition, our model can predict the viral load at ART initiation time, providing another critical public health measure in addition to CD4 counts \citep{braunstein2016}.



We plan to incorporate other variables collected in the HIV surveillance registry as a next step. Including time-invariant covariates such as age at diagnosis and gender in the segmented NLME models is straightforward but with increased computing burden since the M-step won't be in closed-form. Handling time-varying measures such as CD4 (another primary biomarker reported in the HIV surveillance data) is much more involved. As ART initiation would also induce trajectory change to CD4, a practical approach is to expand the current model into a bivariate model to jointly model viral load and CD4. The asynchronous reports between viral load and CD4 generate missing data on both longitudinal data, and handling such missing data in our context is non-trivial. However, we feel such expansion will benefit the estimation accuracy with the enriched within-subject measures and by borrowing information from the correlated biomarkers.

StEM substantially improves the computation efficiency over the MCEM method for the problem when there is no existing analytic form for the E-step. Using an independent sample from rejection sampling, we achieve convergence within 2000 iterations for the size of our problem. Due to the high dimension of the missing data structure, a Gibbs sampler has to be embedded within each StEM iteration. Even though computer power has been increasing tremendously, it is wise to keep the sample size and the number of simulations manageable. Hence, we shall continue in research for even more computationally efficient methods. One possibility is to use crude approximations, e.g., Metropolis sampling \citep{delyon1999}, at a burn-in period and gradually increase the accuracy of the approximation to the proper distribution.

\section*{Appendix: Multivariate Rejection Sampling Algorithm} {Sampling from the full conditional of the random effects can be accomplished by a multivariate rejection algorithm. For univariate density functions are log-concave in the
appropriate parameters, the adaptive rejection algorithm of \cite{gw1992} may be used, as in \cite{ibra1999}. However, for segmented-NLME models, some densities may not be
log-concave, and some are multivariate. In such cases, the multivariate rejection sampling method \cite{gewe1996} may be used to obtain the desired
samples. \cite{bh1999} discussed such a method in the context
of complete data generalized linear mixed models, which can be extended to
our models as follows.

Consider sampling from $f(\bb _{i}|\by_{i},\tau_{i},\ba_{i};\btheta^{(k)})$. Let 
$f^{*}(\bb_{i}) = f(\bb_{i};\bthetat) \cdot f(\by_{i}|\tau_{i},\ba_{i},\bb_{i};\btheta^{(k)})$

and $\xi
= \sup_{\mathbf{u}}\{f^{*}(\mathbf{u})\}$. A random sample
from $f(\bb _{i}|\by_{i},\tau_{i},\ba_{i};\btheta^{(k)})$
 can be obtained as follows:
 \begin{description}
 \item{Step 1:} sample $\bb_{i}^{*}$
 from $f(\bb_{i}; \bthetat)$, and independently,
 sample $w$ from the uniform(0,1) distribution; 
 \item{Step 2:} if $w \leq
 f^{*}(\bb_{i}^{*})/\xi$, then
accept $\bb_{i}^{*}$ as a sample point from $f(\bb _{i}|\by_{i},\tau_{i},\ba_{i};\btheta^{(k)})$, otherwise, go back to step 1 and continue. 
\end{description}
}
\section*{Acknowledgements}

The work is supported by NIH grant R21AI147933 and the CUNY Institute for Implementation Science in Population Health. It is also partially supported by the City University of New York High-Performance Computing
Center, College of Staten Island, funded in part by the City and State of New York, City University
of New York Research Foundation and National Science Foundation grants CNS-0958379, CNS-0855217, and
ACI-112611. 

\newpage
\pagestyle{empty}

\bibliographystyle{Chicago}

\bibliography{mybib}

\newpage

\begin{figure}[ht]
\begin{center}
\centering
\includegraphics[scale=0.75]{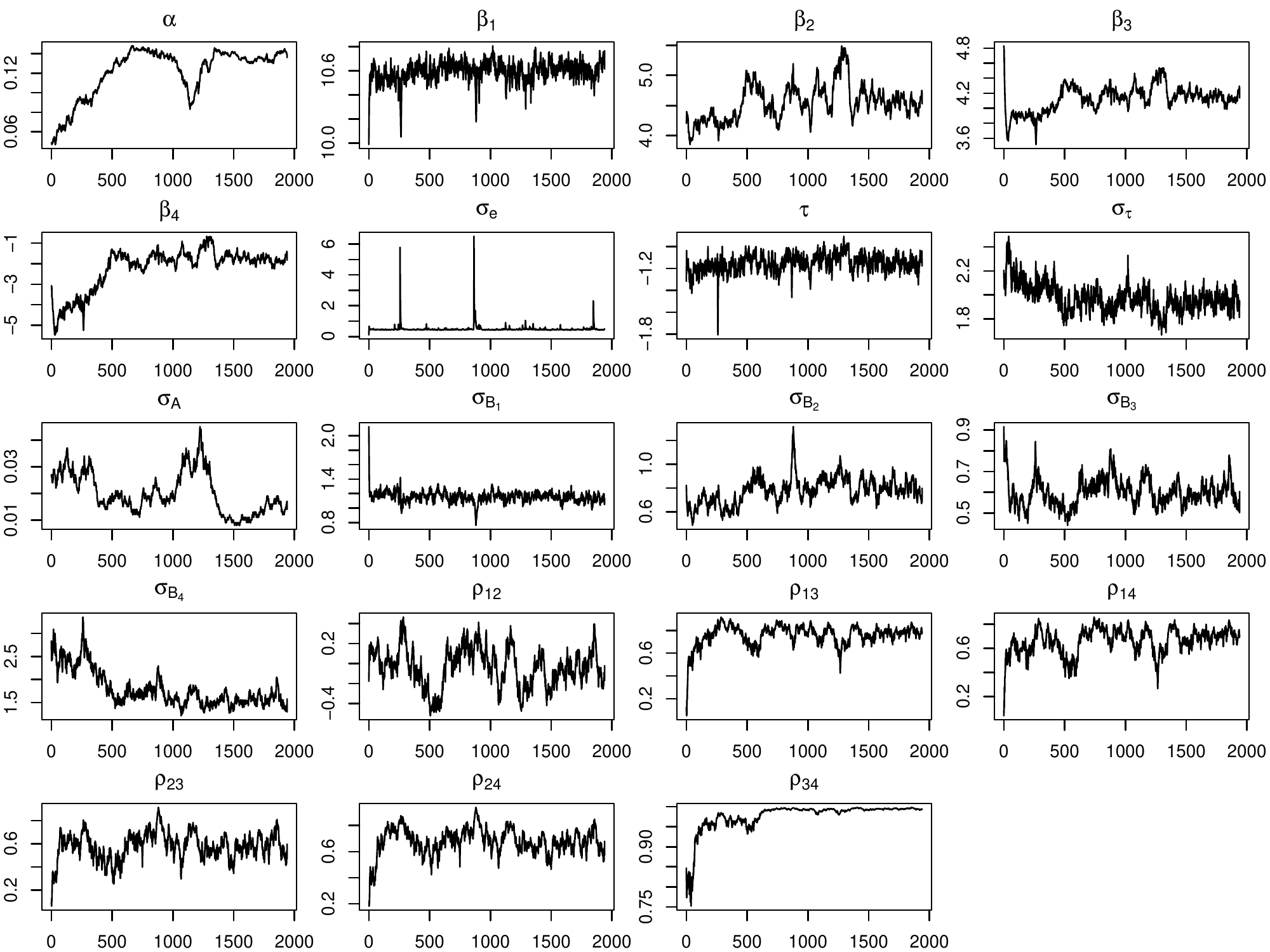}
\caption[]{\label{fig:Fig1} Trace plots of all parameters in the segmented-NLME model for the HIV Registry data. Results from 2000 iterations are displayed although for this sample, the convergence is arrived at the 1940$^{th}$ iteration. The parameters $\sigma_{B_{1}}$, $\sigma_{B_{2}}$, $\sigma_{B_{3}}$, $\sigma_{B_{4}}$ are the standard deviation of random effects $\bb_{i}$, $\rho_{12}$, $\rho_{13}$, $\rho_{14}$, $\rho_{23}$, and $\rho_{34}$ are the corresponding correlation parameters. }
\end{center}
\end{figure}

{\footnotesize
\begin{center}
\begin{table}
\caption{\label{Table1} Estimates in fitting NYC HIV surveillance registry data to the {\sl log1plus$^{*}$} and StEM algorithm}
\centering
\begin{tabular} {llccccccc}
\toprule
Method & Parameter          & $\alpha$ & $\beta_{1}$ &  $\beta_{2}$ & $\beta_{3}$ & $\beta_{4}$ & $\tau$\\
\hline
{\sl $log1plus^{*}$} & est           & 0.05         & 9.99           &  4.20         & 4.83 & -3.07 & -1.32 \\
     & se            & 0.01         &  0.10           &  0.07         & 0.05 &  0.25 &  0.11 \\
 StEM & est           & 0.14         & 10.72           &  4.59         & 4.21 & -1.64 & -1.15 \\
     & se            & 0.03         &  0.12           &  0.21         & 0.08 &  0.23 &  0.17 \\    
\bottomrule
\multicolumn{8}{l}{ Estimates of the variance components:}\\
 \multicolumn{8}{l}{  $log1plus^{*}: \sigma_{A}=0.03$, $\sigma_{\tau}=2.21$, $\sigma_{e}=0.10$,}\\
 \multicolumn{8}{l}{ \quad \quad \quad \quad \quad $B=\begin{pmatrix} 2.13^{2} & -0.31 & 0.24 & 0.24\\
                                              & 0.82^{2} & 0.07 & 0.43\\
                                              &   & 0.92^{2} & 2.20\\
                                              &   &   & 2.84^{2}\\ \end{pmatrix}$}\\
 \multicolumn{8}{l}{  $\text{StEM: } \sigma_{A}=0.02$, $\sigma_{\tau}=1.97$, $\sigma_{e}=0.46$,}\\                                         
 \multicolumn{8}{l}{ \quad \quad \quad \quad \quad  $B=\begin{pmatrix} 1.10^{2} & -0.12 & 0.45 & 1.05\\
                                              & 0.75^{2} & 0.21 & 0.66\\
                                              &   & 0.54^{2} & 0.76\\
                                              &   &   & 1.41^{2}\\ \end{pmatrix}$}\\
\end{tabular}
\end{table}
\end{center}          
}

\begin{figure}[ht]
\begin{center}
\centering
\includegraphics[scale=0.95]{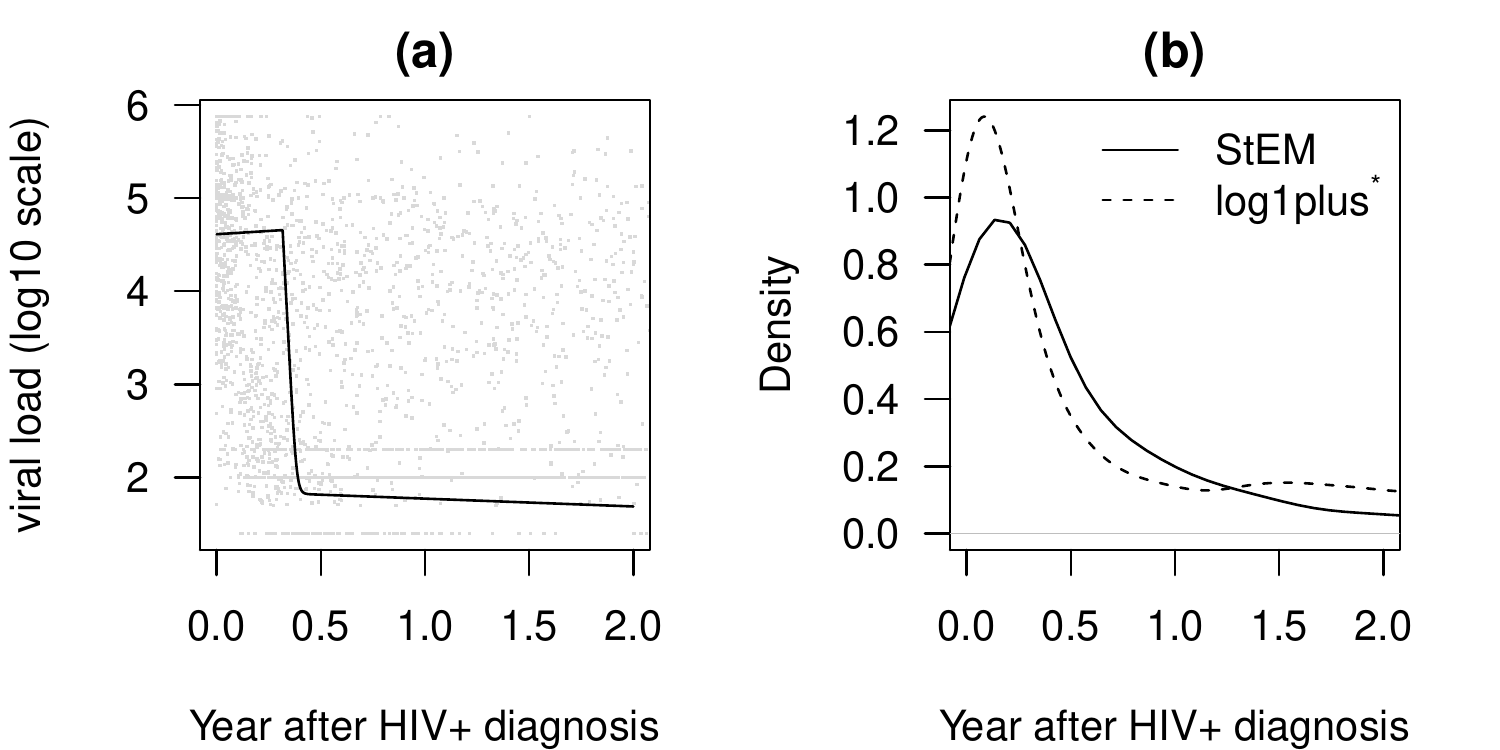}
\caption[]{\label{fig:Fig2} (a) Population level HIV viral load trajectory estimated by the segmented-NLME algorithm; (b) Density plots for the distribution of ART initiation time, from the StEM model-based estimation and from the empirical {\sl log1plus$^{*}$} algorithm. }
\end{center}
\end{figure}
        
\begin{figure}[ht]
\begin{center}
\centering
\includegraphics[scale=0.75]{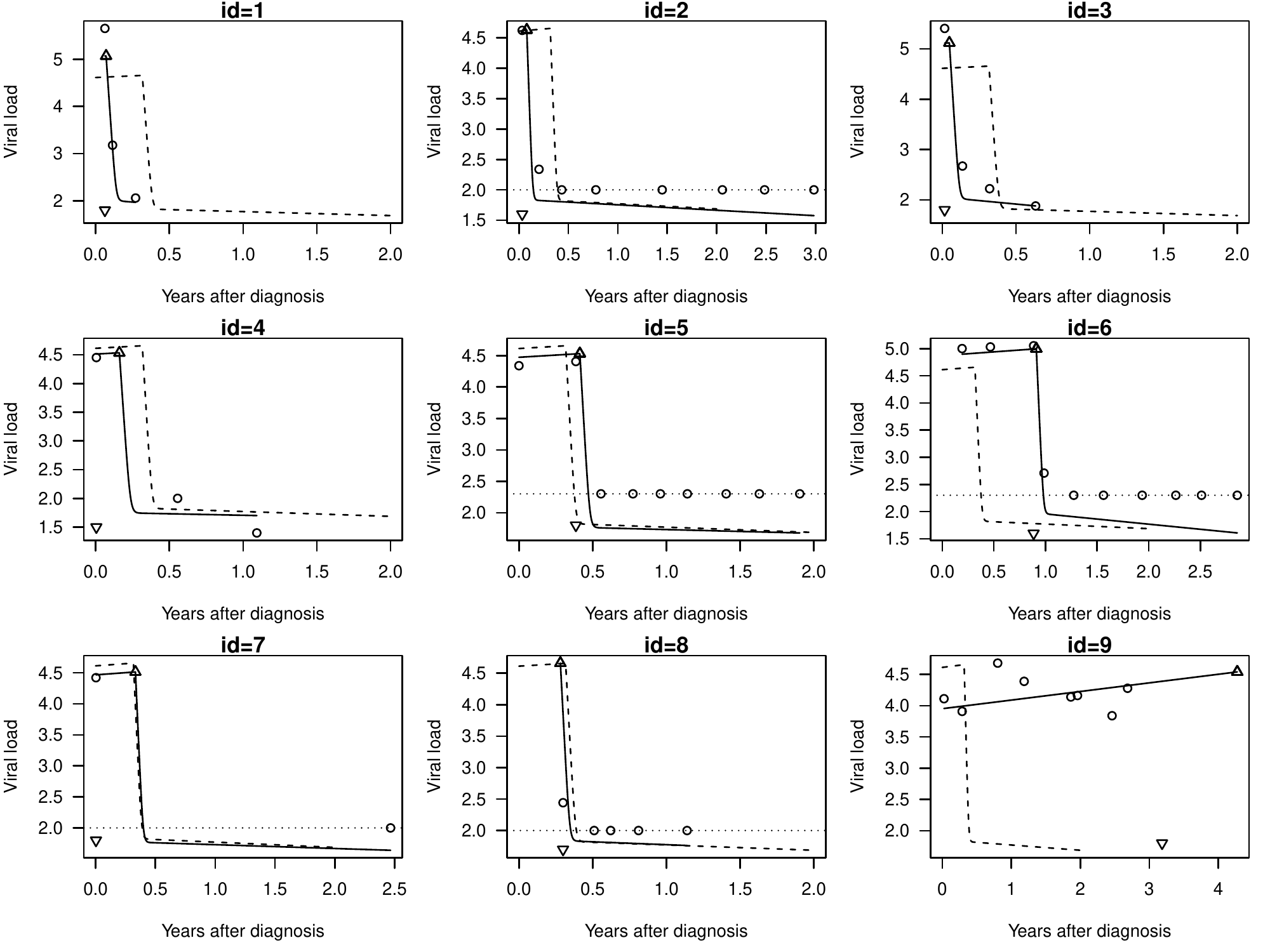}
\caption[]{\label{fig:Fig3} Individual trajectory prediction plots. The open circles ($\circ$) are the observed viral loads, the solid line is the predicted trajectory, and the dashed line is the population trajectory. The detection limit is displayed by dotted line where the viral load is initially imputed by half of the limit when left-censoring occurs (see id=2, 5, 6, 7 and 8 where the detect limit is at 100 units for id 2, 7 and 8 while for id 5 and 6, the threshold is at 200 units). A triangle ({\scriptsize $\triangle$}) is used to display the predicted change point and the viral load at the ART initiation time. A down-pointing triangle ({\small $\triangledown$}) is used to indicate the change point time obtained from the {\sl log1plus$^{*}$ algorithm.} }
\end{center}
\end{figure}

\begin{center}
\begin{table}
\caption{\label{Table2}Simulation results on the performance of the {\sl log1plus$^{*}$} and StEM algorithm}
\centering
\scriptsize
\begin{tabular} {llrrrrrr}
\toprule
  & $\theta$  &   ${\alpha}$ & ${\beta}_{1}$ & $\beta_{2}$ &  $\beta_{3}$ & $\beta_{4}$ & ${\tau}$  \\
 \addlinespace 
  Method      &     True           &   {0.14} & {10.72} & {4.59} &  {4.21}  & {-1.64} & -1.15 \\
\hline
\addlinespace
\multicolumn{2}{c}{ } & \multicolumn{6}{c}{{\tiny {\bf 0\% left-censored}}}\\
\addlinespace                    
  $log1plus^{*}$& Est       & 0.01 &  9.61 & 4.36 & 4.00 &  -2.26     &  -1.16  \\
    & SE        & 0.03 &  0.13 &  0.13 &  0.04 &  0.14         &  0.08    \\
    & SD        & 0.01 &  0.01 &  0.06 &  0.02 &  0.10        &  0.01   \\
    & MSE       & 0.02 &  1.25 & 0.07 &  0.05 & 0.42           & 0.01    \\              
    & Bias\%       & -90.00 &  -10.35 & -5.00 &  -4.99 &  37.98      &  0.61 \\
\addlinespace        
  StEM& Est       & 0.11 &  10.69 & 4.74 & 4.22 &  -1.43                          &  -1.14  \\
    & SE        & 0.03    &  0.14    &  0.28    &  0.07    &  0.12            &  0.16   \\
    & SD        & 0.03    &  0.10    &  0.26    &  0.06    &  0.15            &  0.16  \\
    & MSE       & 0.01    &  0.21    & 0.16     & 0.02     & 0.07             & 0.05   \\              
    & Bias\%       & -24.24    &  -4.05    & 2.10     &  1.65    & -11.56             &  -0.37   \\
\addlinespace
\multicolumn{2}{c}{ } & \multicolumn{6}{c}{{\tiny {\bf 30\% left-censored}}}\\
\addlinespace                    
 $log1plus^{*}$ & Est       & 0.08 & 10.51 & 3.87 & 4.62 &  -2.69     &  -2.44    \\
    & SE        & 0.05 &  0.07 &  0.10 &  0.04 &  0.21         &  0.09    \\
    & SD        & 0.06 &  0.07 &  0.07 &  0.04 &  0.33        &  0.10    \\
    & MSE    & 0.01 & 0.05 & 0.53 &  0.17 &  1.26      &  1.68   \\
    & Bias\%       & -41.46 &  -1.97 & -15.60 &  9.69 & 64.19           & 112.22  \\              
\addlinespace        
  StEM& Est       & 0.15 &  10.72 & 4.67 & 4.24 &  -1.64                          &  -1.20    \\
    & SE        & 0.03    &  0.05    &  0.21    &  0.03    &  0.10            &  0.09    \\
    & SD        & 0.04    &  0.06    &  0.22    &  0.05    & 0.22             &  0.06    \\
    & MSE       & 0.01    &  0.01    &  0.10     &  0.01    & 0.06          &  0.01   \\
    & Bias\%      & 5.56    &  0.04    & 1.83     & 0.83     & -0.04             & 4.33   \\                 
\addlinespace
\multicolumn{2}{c}{ } & \multicolumn{6}{c}{{\tiny {\bf 50\% left-censored}}}\\
\addlinespace                    
 $log1plus^{*}$ & Est       & 0.14 &  10.57 & 3.89 & 5.00 &  -3.89     &  -2.48    \\
    & SE        & 0.05 &  0.07 &  0.10 &  0.03 &  0.28         &  0.09    \\
    & SD        & 0.05 &  0.07 &  0.07 &  0.03 &  0.46        &  0.09    \\
    & MSE    & 0.01 & 0.03 & 0.50 &  0.63 &  5.37      &  1.78   \\
    & Bias\%       & -0.05 &  -1.41 & -15.21 &  18.88 & 137.38           & 115.46  \\              
\addlinespace        
  StEM& Est       & 0.14 &  10.72 & 4.73 & 4.33 &  -1.68                          &  -1.19    \\
    & SE        & 0.03    &  0.05    &  0.23    &  0.04    &  0.11            &  0.09    \\
    & SD        & 0.05    &  0.06    &  0.30    &  0.08    & 0.40             &  0.07    \\
    & MSE       & 0.01    &  0.01    &  0.16     &  0.02    & 0.17          &  0.01   \\
    & Bias\%      & 1.05    &  0.02    & 3.04     & 2.88     & 2.37             & 3.37   \\                 
\addlinespace
\bottomrule
\multicolumn{8}{l}{MSE = $\frac{1}{S}\sum_{s=1}^{S} 100 \times \frac{\sqrt{(\hat{\theta}^{(s)} - \theta^{\text{true}})^2 + \text{SE}(\hat{\theta}^{(s)})}}{|\theta^{true}|}$   }\\
\multicolumn{8}{l}{Bias\% = $\frac{1}{S} \sum_{s=1}^{S} 100 \times (\hat{\theta}^{(s)} - \theta^{\text{true}})/\theta^{\text{true}}$   }\\
\end{tabular}
\end{table}
\end{center}

\end{document}